\documentclass[aps,prx,showpacs,floatfix,twocolumn,superscriptaddress,longbibliography]{revtex4-2}

\usepackage{float}
\usepackage{color}
\usepackage{bm}
\usepackage{hyperref}
\usepackage{todonotes}
\usepackage{verbatim}
\usepackage{soul}
\usepackage{glossaries}
\usepackage{sidecap}
\usepackage{hyperref}
\hypersetup{
    colorlinks=true,
    linkcolor=blue,
    filecolor=magenta,
    urlcolor=red,
    citecolor=blue,
}

\def\Q{\ensuremath{\mathbf{Q}}}
\def\QCO{\ensuremath{\mathbf{Q}_{\mathrm{CO}}}}
\def\DEd{\ensuremath{\Delta\epsilon_d}}

\def\La438{La$_{4}$Ni$_{3}$O$_{8}$}
\def\Pr438{Pr$_{4}$Ni$_{3}$O$_{8}$}
\def\Ni112{\textit{R}NiO$_2$}

\newacronym{CO}{CO}{charge order}
\newacronym{JDOS}{JDOS}{joint density of states}
\newacronym{BS}{BS}{bond-stretching}
\newacronym{RIXS}{RIXS}{resonant inelastic x-ray scattering}
\newacronym{REXS}{REXS}{resonant elastic x-ray scattering}
\newacronym{XAS}{XAS}{x-ray absorption spectrum}
\newacronym{EELS}{EELS}{electron energy loss spectroscopy}
\newacronym{EPC}{EPC}{electron-phonon coupling}
\newacronym{CDW}{CDW}{charge density wave}
\newacronym{SDW}{SDW}{spin density wave}
\newacronym{FWHM}{FWHM}{full-width at half-maximum}
\newacronym{INS}{INS}{inelastic neutron scattering}
\newacronym{DFT}{DFT}{density functional theory}
\newacronym{GGA}{GGA}{generalized gradient approximation}
\newacronym{UHB}{UHB}{upper Hubbard band}
\newacronym{ZSA}{ZSA}{Zaanen-Sawatzky-Allen}
\newacronym{ZRS}{ZRS}{Zhang-Rice singlet}
\newacronym{ED}{ED}{exact diagonalization}
\newacronym{CEF}{CEF}{crystal electric field}
\newacronym{2D}{2D}{two-dimensional}
\newacronym{TM}{TM}{transition-metal}
\newacronym{LDA}{LDA}{local density approximation}
\newacronym{DMFT}{DMFT}{dynamical mean field theory}

\begin{document}

\title{Electronic character of charge order in square planar low valence nickelates}

\author{Y. Shen}\email[]{yshen@bnl.gov}
\affiliation{Condensed Matter Physics and Materials Science Department, Brookhaven National Laboratory, Upton, New York 11973, USA}

\author{J. Sears}
\affiliation{Condensed Matter Physics and Materials Science Department, Brookhaven National Laboratory, Upton, New York 11973, USA}

\author{G. Fabbris}
\affiliation{Advanced Photon Source, Argonne National Laboratory, Lemont, Illinois 60439, USA}

\author{J. Li}
\author{J. Pelliciari}
\affiliation{National Synchrotron Light Source II, Brookhaven National Laboratory, Upton, New York 11973, USA}

\author{M. Mitrano}
\affiliation{Department of Physics, Harvard University, Cambridge, Massachusetts 02138, USA}

\author{W. He}
\affiliation{Condensed Matter Physics and Materials Science Department, Brookhaven National Laboratory, Upton, New York 11973, USA}

\author{Junjie Zhang}
\affiliation{Materials Science Division, Argonne National Laboratory, Lemont, Illinois 60439, USA}
\affiliation{Institute of Crystal Materials, Shandong University, Jinan, Shandong 250100, China}

\author{J. F. Mitchell}
\affiliation{Materials Science Division, Argonne National Laboratory, Lemont, Illinois 60439, USA}

\author{V. Bisogni}
\affiliation{National Synchrotron Light Source II, Brookhaven National Laboratory, Upton, New York 11973, USA}

\author{M. R. Norman}
\affiliation{Materials Science Division, Argonne National Laboratory, Lemont, Illinois 60439, USA}

\author{S. Johnston}
\affiliation{Department of Physics and Astronomy, The University of Tennessee, Knoxville, Tennessee 37966, USA}
\affiliation{Institute of Advanced Materials and Manufacturing, The University of Tennessee, Knoxville, Tennessee 37996, USA\looseness=-1}

\author{M. P. M. Dean}\email[]{mdean@bnl.gov}
\affiliation{Condensed Matter Physics and Materials Science Department, Brookhaven National Laboratory, Upton, New York 11973, USA}

\date{\today}

\begin{abstract}
Charge order is a central feature of the physics of cuprate superconductors and is known to arise from a modulation of holes with primarily oxygen character. Low-valence nickelate superconductors also host charge order, but the electronic character of this symmetry breaking is unsettled. Here, using resonant inelastic x-ray scattering at the Ni $L_2$-edge, we identify intertwined involvements of Ni $3d_{x^2-y^2}$, $3d_{3z^2-r^2}$, and O $2p_{\sigma}$ orbitals in the formation of diagonal charge order in an overdoped low-valence nickelate \La438{}. The Ni $3d_{x^2-y^2}$ orbitals, strongly hybridized with planar O $2p_{\sigma}$, largely shape the spatial charge distribution and lead to Ni site-centered charge order. The $3d_{3z^2-r^2}$ orbitals play a small, but non-negligible role in the charge order as they hybridize with the rare-earth $5d$ orbitals. Our results reveal that the low-energy physics and ground-state character of these nickelates are more complex than those in cuprates.
\end{abstract}

\maketitle

\section{Introduction}

One of the common threads linking different classes of unconventional superconductors is their propensity to host proximate competing orders such as charge and spin stripes \cite{Norman2011challenge,Scalapino2012common}. For example, the cuprate superconductors exhibit diagonal (with respect to the Cu-O bonds) spin stripes when underdoped  \cite{Yamada1998doping,Wakimoto2001Diagonal,Dunsiger2008Incommensurate}, while Cu-O bond oriented (parallel) charge order dominates the rest of the phase diagram \cite{Arpaia2019Dynamical,Miao2021Charge}. The detection of superconductivity and charge order in the square-planar low-valence family of nickelates therefore presents a fascinating opportunity to study the degree of similarity between different unconventional superconducting families \cite{Li2020superconduting,Zeng2020phase,Osada2020phase,Osada2021nickelates,Zeng2022Superconductivity,Pan2021super,Zhang2016stacked,Rossi2021Broken,Tam2021Charge,Krieger2021Charge}. Intriguingly, different nickelates within the structural series of $R_{n+1}$Ni$_{n}$O$_{2n+2}$ (\textit{R} stands for a rare earth and $n$ is the number of neighboring NiO$_2$ layers) also host different charge ordered phases. Underdoped materials with $n=\infty$ and $R=$~La, Nd exhibit parallel charge order \cite{Rossi2021Broken,Tam2021Charge,Krieger2021Charge}, whereas $n=3$ material \La438{}, which is effectively $1/3$ overdoped, manifests diagonal charge order \cite{Zhang2016stacked}. Many researchers have emphasized that charge order plays an important role in the physics of cuprates \cite{Tranquada2013stripes,Comin2016review,Frano2020charge,Tranquada2021SC}. In particular, there is good evidence showing that charge/spin order is a fundamental feature of minimal Hubbard model descriptions of the cuprates \cite{Huang2017numerical, Zheng2017stripe, Mai2022stripes}. Some researchers have suggested that charge and spin order can intertwine with superconductivity to form pair density waves \cite{Li2007two, Berg2007dynamical}, or that dynamic charge/spin fluctuations might promote superconductivity \cite{Emery1997spin, Kivelson1998electronic, Fradkin2015Theory}. Others have associated charge order fluctuations with the anomalous ``strange metal'' electronic transport in cuprates \cite{Castellani1995singular}. Understanding the electronic states involved in charge order formation is a prerequisite to testing all these scenarios in low-valence nickelates and is also important more generally for understanding charge order as a prevalent feature of correlated quantum materials.

\begin{figure*}
\includegraphics{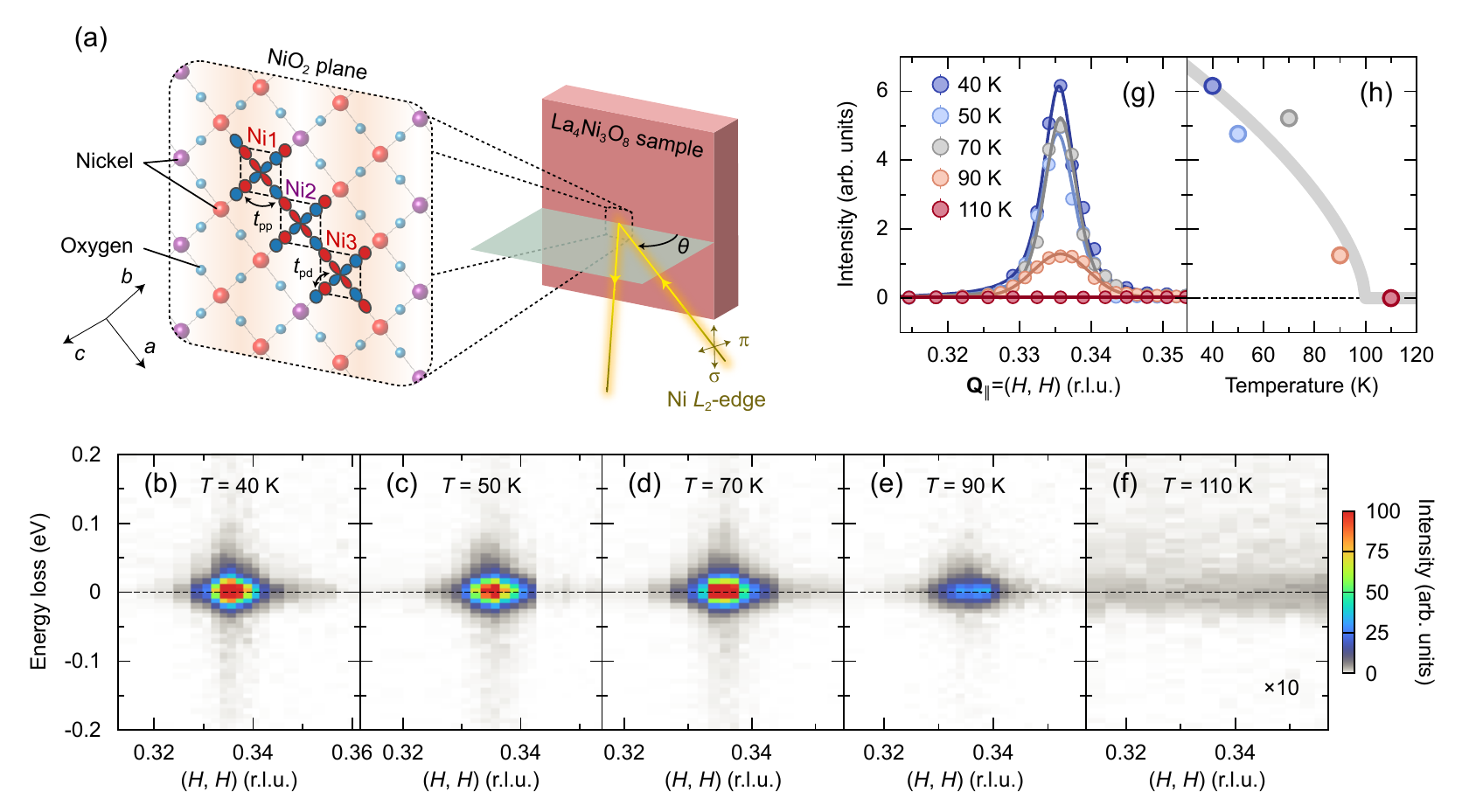}
\caption{Charge order transition in \La438{}. (a) Schematic of the Ni $L_2$-edge \gls*{RIXS} experimental setup. A single NiO$_2$ layer is presented with stripes running vertically. A Ni$_3$O$_{10}$ cluster composed of Ni $3d_{x^2-y^2}$ and planar O $2p_{\sigma}$ orbitals is embedded in it, tracing the charge order motif, in which hole poor Ni1 and Ni3 sites, shown in red, flank the hole rich Ni2 site depicted in purple. (b)--(f) \gls*{RIXS} intensity maps with $\sigma$ polarized incident photons at the indicated temperatures obtained by changing the in-plane sample angle $\theta$. (g) Quasi-elastic-line amplitudes extracted from the data presented in (b)--(f) as a function of in-plane momentum transfer in reciprocal lattice units (r.l.u.). The solid lines are fitting curves with pseudo-Voigt profiles. (h) Temperature dependence of the fitted peak amplitudes. The bold gray line is a guide to the eye.}
\label{fig:schematic}
\end{figure*}

Here, we use Ni $L_2$-edge \gls*{RIXS} to determine the electronic character of the charge order in \La438{}. We find that both the Ni $3d_{x^2-y^2}$ and $3d_{3z^2-r^2}$ orbitals are involved in charge order formation. The former contributes most of the charge modulation while the latter dominates the \gls*{RIXS} spectra in the post-edge regime and so plays a less important role. As the charge-transfer energy of these nickelates is larger than that of cuprates but comparable to the on-site Coulomb interaction, the holes involved in the charge modulation reside predominately on Ni sites, despite an appreciable amount of holes occupying the O orbitals. Our results indicate that the low-energy electronic structure and charge order of low-valence nickelates is largely shaped by hybridized $3d_{x^2-y^2}$ and planar O $2p_{\sigma}$ orbitals, similar to cuprates, while some differences exist due to the multi-band physics introduced by Ni $3d_{3z^2-r^2}$ orbitals hybridized with rare-earth $5d$ states.

\begin{figure*}
\includegraphics{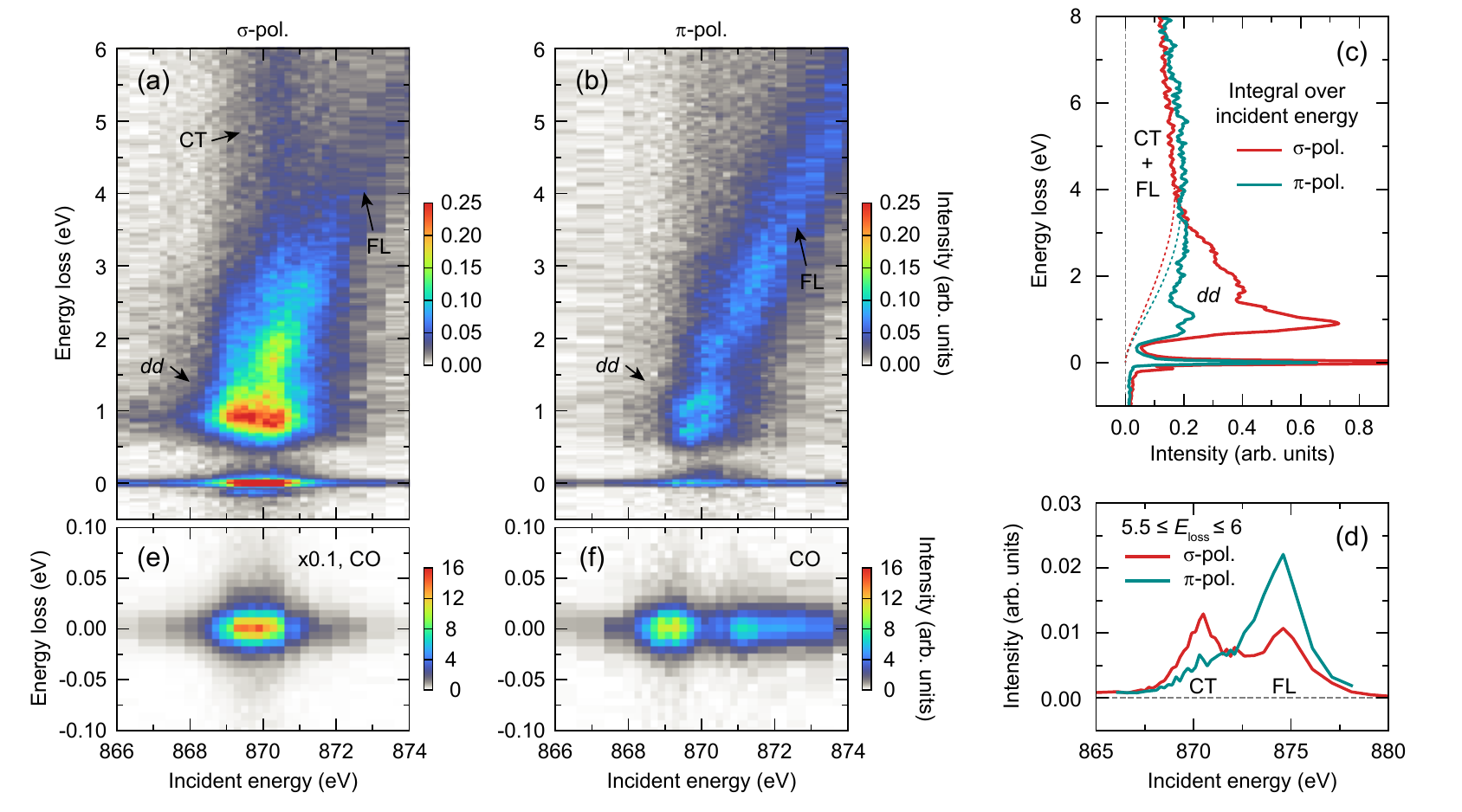}
\caption{\gls*{RIXS} energy maps and the resonant behaviors of the charge order (CO) peak. (a,~b) \gls*{RIXS} intensity maps as a function of incident photon energy with (a) $\sigma$ x-ray polarization in the $ab$ plane of the sample and (b) $\pi$ x-ray polarization approximately parallel to the $c$-axis. Several components can be identified: charge transfer excitations (CT), $dd$ excitations ($dd$) and constant-emission-energy fluorescence (FL). (c) Integral of the \gls*{RIXS} spectra along the incident energy axis. The dashed lines are guides to the eye. (d) Incident energy dependence of the integrated \gls*{RIXS} spectra between 5.5 and 6~eV energy loss. (e,~f) \gls*{RIXS} intensity maps around the quasi-elastic regime with \Q{} fixed at \QCO{}. Note that the intensity in (e) is multiplied by 0.1 for clarity in visualizing the signal.}
\label{fig:Emap}
\end{figure*}

\section{Results}
The \La438{} nickelate samples studied here were prepared by reducing single crystals synthesized via the floating zone method (see the Appendix~\ref{sec:synthesis} for details), and will be indexed in terms of scattering vector $\mathbf{Q}=(2\pi/a, 2\pi/a, 2\pi/c)$ with $a=b=3.97$~\AA{}, $c=26.092$~\AA{}. As the $n=3$ member of the low-valence nickelate family, it possesses a trilayer structure with a nominal $3d^{8+2/3}$ valence. This leads to a 1/3-hole self-doping with respect to the undoped $3d^9$ state, putting it in the overdoped regime of the phase diagram \cite{Zhang2017large,Pan2021super}. It shares the same structural motif as infinite-layer nickelates with square-planar NiO$_2$ layers stacked without apical oxygens, leading to dominant Ni $3d_{x^2-y^2}$ character near the Fermi energy. Although \La438{} has two inequivalent NiO$_2$ layers, they are expected to show similar electronic structure as indicated by theoretical calculations \cite{Karp2020comparative,Nica2020theoretical}, which is further supported by the observation that the same charge order pattern is formed in both layers \cite{Zhang2016stacked}. We study their properties using Ni $L_2$-edge \gls*{RIXS} in order to avoid interference from the La $M_4$-edge, which overlaps the Ni $L_3$-edge  (see the  Appendix~\ref{sec:RIXS} for details). As shown in  Fig.~\ref{fig:schematic}(a), charge order in \La438{} is quasi-two-dimensional in nature and occurs at $\mathbf{Q}_{\parallel}=\mathbf{Q}_{\mathrm{CO}}=(1/3, 1/3)$, where a strong peak is observed in the quasi-elastic region of the \gls*{RIXS} intensity map at 40~K [see Fig.~\ref{fig:schematic}(b)]. The in-plane correlation length is larger than 100~nm, which might be limited by the sample mosaic, suggesting the long range nature of the charge order \cite{Zhang2016stacked}. This charge order peak persists up to 90~K and disappears above 110~K, indicating a transition temperature of around 100~K [see Figs.~\ref{fig:schematic}(c)--(h)], consistent with the reported charge order from hard x-ray diffraction measurements \cite{Zhang2016stacked}. No indication of charge order is apparent in equivalent measurements of metallic \Pr438{} samples prepared in the same way (Supplemental Material Sec.~I \cite{supp}).

\begin{figure}
\includegraphics{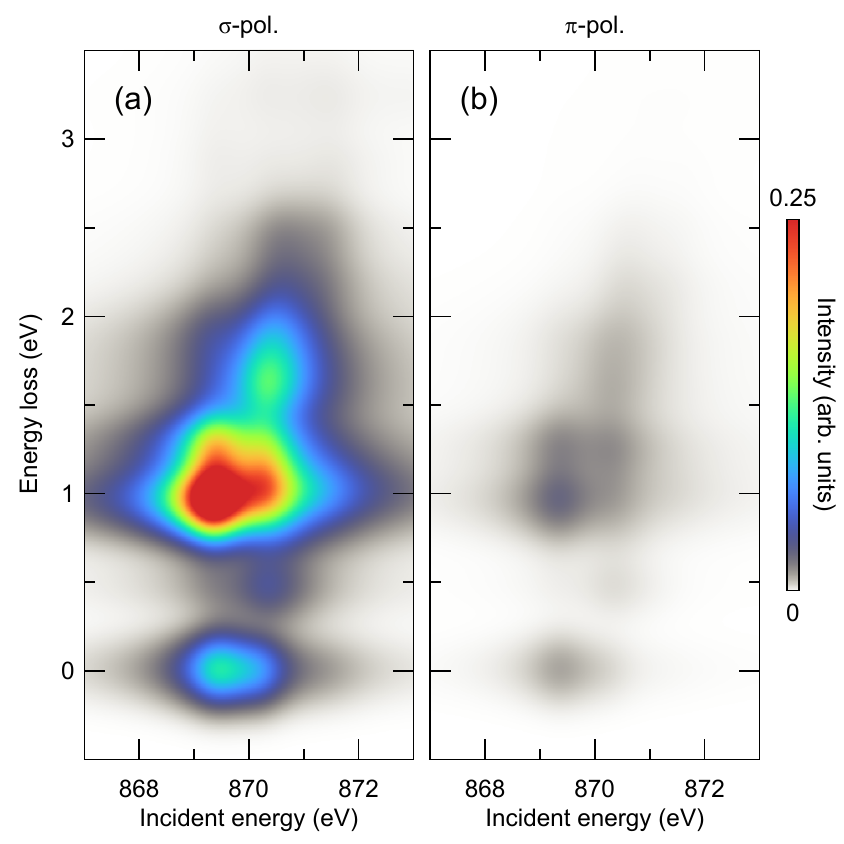}
\caption{Low-energy electronic states in \La438{}. Calculations of the \gls*{RIXS} energy maps at the Ni $L_2$-edge for (a) $\sigma$ and (b) $\pi$ incident x-ray polarization. The calculations reproduce the experimental energy-scale and polarization of the $dd$ excitations evincing an appropriate minimal model for \La438{}.}
\label{fig:EmapCalc}
\end{figure}

We begin by identifying the active electronic states in \La438{} using x-ray spectroscopy. Figure~\ref{fig:Emap}(a) and \ref{fig:Emap}(b) show the $L_2$-edge \gls*{RIXS} energy maps taken with $\sigma$ x-ray polarization in the $ab$-plane and $\pi$ x-ray polarization approximately parallel to the $c$-axis, respectively. The \gls*{RIXS} maps mainly comprise $dd$ and charge-transfer excitations that are predominantly localized and resonate at the Ni $L_2$-edge, and diagonal fluorescence features (Supplemental Material Sec.~II \cite{supp}). To distinguish among these contributions, we integrated the \gls*{RIXS} spectra along the incident energy axis and show the result in Fig.~\ref{fig:Emap}(c). With $\sigma$ polarization, the spectra above 4~eV energy loss are dominated by mostly featureless fluorescence originating from particle-hole excitations that can be understood from an itinerant framework involving transitions from extended electronic bands spanning many unit cells \cite{Fabbris2017doping}. Charge transfer excitations are also visible above 4~eV but only at resonance. Below 4~eV, prominent $dd$ excitations emerge that dominate over the featureless fluorescence (dashed lines). With $\pi$ polarization, the fluorescence contributes most of the spectral weight and the $dd$ excitations are much weaker. The strong dichroism of $dd$ excitations reflects the dominant Ni $3d_{x^2-y^2}$ orbital character near the Fermi energy in low-valence nickelates.

To further distinguish between charge-transfer excitations and fluorescence, we inspect the \gls*{RIXS} spectra between 5.5 and 6~eV energy loss, well above the $dd$ excitation threshold. As shown in Fig.~\ref{fig:Emap}(d), the charge-transfer excitations and fluorescence are separated in the incident energy axis, with the former stronger in the $\sigma$ polarization channel, indicating appreciable $d_{x^2-y^2}$-$p_{\sigma}$ hybridization where $p_{\sigma}$ indicates O orbitals that are parallel to the Ni-O bonds. In contrast, the fluorescence is stronger in the $\pi$ polarization channel, suggesting that states involving Ni $3d_{3z^2-r^2}$ orbitals dominate the fluorescence for a broad range of energy losses above $\sim$3~eV. The broadness of these states is in contrast with cuprates, and suggests that although the Ni $3d_{3z^2-r^2}$ orbitals are mostly occupied and localized, their unoccupied components are hybridized with the rare earth $5d$ orbitals and thus contribute to dispersive states. This conclusion is consistent with \gls*{DFT}+\gls*{DMFT} calculations \cite{Karp2020comparative}, as well as \gls*{RIXS} simulations for $R$NiO$_2$ that studied the effect of switching on and off the rare-earth hybridization \cite{Higashi2021core}. Meanwhile, the Ni $3d_{x^2-y^2}$ orbitals exhibit less hybridization with the rare earth $5d$ orbitals and are more localized. Here, since we are measuring at the Ni $L$ edge and the Ni $t_{2g}$ orbitals are expected to lie well below the Fermi energy, we only consider Ni $e_g$ orbitals \cite{supp}.

Based on the resonant behavior of the different states identified, we now examine how the $3d_{x^2-y^2}$ and $3d_{3z^2-r^2}$ orbitals participate in the charge order. Figure~\ref{fig:Emap}(e) and \ref{fig:Emap}(f) show the \gls*{RIXS} energy maps around the quasi-elastic regime at \QCO{}, i.e.\ the \gls*{REXS} signals. The peak intensity strongly resonates at the Ni $L_2$-edge in the $\sigma$ polarization channel [see Fig.~\ref{fig:Emap}(e)], confirming that the $(1/3,1/3)$ Bragg peak in \La438{} involves a charge modulation and is not purely structural. Surprisingly, the charge order peak in the $\pi$ polarization channel, although much weaker, resonates at the pre- and post-edge regimes but not at the main edge [see Fig.~\ref{fig:Emap}(f)], distinct from that in cuprates \cite{Abbamonte2005Spatially,Peng2018Reentrant,Li2020Multiorbital}. First, this observation indicates that both the $3d_{x^2-y^2}$ and $3d_{3z^2-r^2}$ orbitals are involved in charge order formation with the latter much less prominent. Second, the charge order peak in the post-edge regime with $\pi$ polarization suggests that the states far above the Fermi energy also show charge modulation, which is mostly contributed by $3d_{3z^2-r^2}$ orbitals. Considering that the $3d_{3z^2-r^2}$ density of states in the post-edge regime is likely caused by hybridization with the rare-earth $5d$ orbitals, this indicates potential involvement of rare-earth orbitals in the charge order formation. Similarly, the weak pre-edge charge order peak with $\pi$ polarization indicates that the $3d_{3z^2-r^2}$ density of states near the Fermi energy is nonzero but small.

Having established the involvement of Ni orbitals in the charge order formation, now we look at the role of oxygen states. To do this, we use \gls*{ED} methods which allow us to solve the resonant cross-section and break down the contributions from different states. Since the charge order is commensurate with a period of three Ni sites and there is a strong hybridization between the Ni and O orbitals, the smallest cluster one can use to describe the charge-ordered state involves three Ni-O plaquettes, which we label 1, 2, \& 3. We choose a bond-oriented cluster, as illustrated in Fig.~\ref{fig:schematic}(a), given that the Ni-O hopping dominates the kinetic energy. In order to compute \gls*{REXS} we use the atomic scattering factors from the cluster and add these amplitudes to simulate an effective two-dimensional NiO$_2$ plane as shown in Fig.~\ref{fig:schematic}(a). The appropriate parameters for this cluster, and in particular the charge-transfer energy $\Delta=5.6$~eV and the on-site Coulomb repulsion $U_{\mathrm{dd}}=6.5$~eV, have been empirically determined by prior x-ray measurements of this material at the O $K$-edge \cite{Shen2022Role}. We use open boundary conditions and construct the Hamiltonian in the hole language  (see the Appendix~\ref{sec:ED} for details). Four holes are introduced to the cluster, which is appropriate for the $d^{9-1/3}$ electronic configuration of \La438{}. Without any additional constraints, the holes will be evenly distributed among different NiO$_4$ plaquettes with minimal charge disproportionation and no symmetry breaking is expected. To realize the charge order observed in \La438{}, we manually introduce a potential difference \cite{Achkar2013resonant}, \DEd{}, for different Ni sites by lowering the orbital energies of Ni2 by $2\Delta\epsilon_d/3$ and raising those of Ni1 and Ni3 by $\Delta\epsilon_d/3$. Based on the similar magnetic exchange of charge ordered \La438{} and metallic \Pr438{} \cite{Lin2021strong}, \DEd{} must be significantly smaller than the charge-transfer energy. Thus, we choose it to be $\Delta\epsilon_d=0.8$~eV while noting that apart from modulating the intensity of the charge order peak, the results are similar provided \DEd{} is not made unfeasibly large (Supplemental Material Fig.~S5 \cite{supp}). This choice leads to a charge disproportionation of $\Delta n=0.32$, which is of a similar order of magnitude as that in cuprates \cite{Abbamonte2005Spatially}. This value is much smaller than the fully disproportionate limit $\Delta n=1$, consistent with \gls*{DFT} calculations that indicate a small charge modulation upon charge ordering in this system \cite{Zhang2017large}. When examining the electronic configuration of the cluster, we find that the ground state is a singlet, and the first excited state is a triplet, which is around 70~meV above the ground state, consistent with the magnetic excitations found in \La438{} \cite{Lin2021strong}.

Figure~\ref{fig:EmapCalc} shows the calculated Ni $L_2$-edge \gls*{RIXS} energy maps with all the Ni $3d$ and O $2p$ orbitals included, which qualitatively reproduce the localized $dd$ excitations observed experimentally. Note that the small cluster size means that we can only capture a limited number of discrete states. For this reason, fluorescence features are not fully captured, which would require a continuous distribution of states. This can be seen more clearly in the $\pi$ polarization channel where the fluorescence dominates the spectra in experimental data [see Fig.~\ref{fig:Emap}(b)] but only the weak $dd$ excitations are present in our cluster calculations [see Fig.~\ref{fig:EmapCalc}(b)].

\begin{figure}
\includegraphics{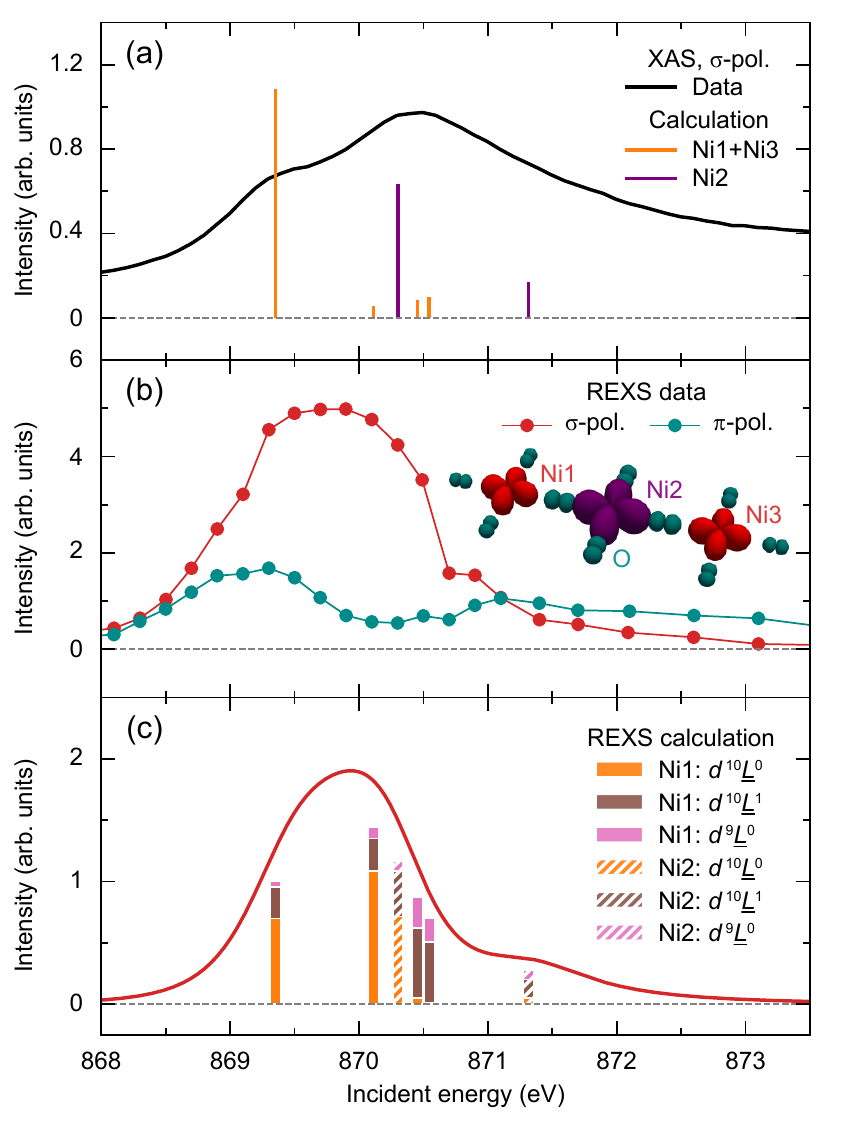}
\caption{Electronic character of charge order. (a) \acrfull*{XAS} data at the Ni $L_2$ edge in the $\sigma$ polarization channel along with the calculation results with $\Delta\epsilon_d=0.8$~eV. Note that Ni1 and Ni3 are symmetry-related. (b) Fitted peak amplitudes of the quasi-elastic intensities presented in Fig.~\ref{fig:Emap}(e)\&(f), representing the resonant behaviors of the charge order peak. Inset is a schematic of the electronic character of the charge order showing a dominant modulation of Ni orbitals along with an appreciable modulation of the oxygen orbitals. (c) Simulation of the incident energy dependence of the charge order peak intensity with $\sigma$ incident polarization and $\Delta\epsilon_d=0.8$~eV. The vertical bars are weights of different configurations of the \gls*{RIXS} intermediate states, the total height of which is normalized according to the simulated charge order peak intensity of each state. The accurate simulation of the Ni $3d$ and O $2p$ components of the resonance verifies our model, which is used to extract the electronic character of the charge order illustrated in the inset to panel (b).}
\label{fig:calc}
\end{figure}

Having verified the relevant parameters via the \gls*{RIXS} maps, we computed the \gls*{XAS} and \gls*{REXS} response of \La438{} using a similar \gls*{ED} approach and identical parameters and plot the results in Fig.~\ref{fig:calc} (Supplemental Material Sec.~V \cite{supp}). The charge disproportionation in the cluster implies a \gls*{REXS} response at \QCO{}. The predicted \gls*{REXS} resonance shown in Fig.~\ref{fig:calc}(c) nicely captures the main two peak structure of the experimental \gls*{REXS} resonance shown in Fig.~\ref{fig:calc}(b). The same applies for the \gls*{XAS} as shown in Fig.~\ref{fig:calc}(a). In fact, the lineshape of the resonant profile of the charge order peak is sensitive to the charge-transfer energy, and neither the pure charge-transfer nor Mott-Hubbard scenarios can describe the observed resonant behaviors (Supplemental Material Fig.~S6 \cite{supp}), demonstrating the mixed charge-transfer/Mott-Hubbard characters of charge order in this material. To understand the nature of the two resonant features, we projected the wavefunctions of the \gls*{RIXS} intermediate states onto the Fock basis which specifies the location of the holes. Two main manifolds are seen for each Ni site. The first manifold is primarily attributed to transitions resonant with $d^{10}\underline{L}^0$ states, where $\underline{L}$ stands for ligand holes on the four oxygen $\sigma$ orbitals surrounding the Ni site. The second manifold is mainly resonant with $d^{9}\underline{L}^0$ and $d^{10}\underline{L}^1$ states caused by the doped holes, similar to the cuprates \cite{Chen1992out,Schneider2005Evolution}. With nonzero \DEd{}, the manifolds of different Ni sites split along the incident energy axis, as shown in Fig.~\ref{fig:calc}(c). The successful description of the charge order in \La438{} using our cluster model indicates that about 70\% of the holes participating in the charge modulation are on Ni, with the remaining 30\% on oxygen, as depicted in the inset to Fig.~\ref{fig:calc}(b).

\section{Discussion}
Our Ni-dominant charge order distribution is quite different from cuprates, in which the charge order has dominant oxygen character \cite{Zhang1988effective, Abbamonte2005Spatially}. This difference mainly arises from the larger charge transfer energy in nickelates compared to cuprates. Another difference is that in cuprates, the $3d_{3z^2-r^2}$ orbitals are strongly localized at energies more than 1.5~eV away from the $3d_{x^2-y^2}$ orbitals \cite{Moretti2011energy}, and thus not involved in the low-energy physics. For square-planar nickelates, our analysis of \La438{} indicates that the $3d_{3z^2-r^2}$ density of states, though small, is spread out over an extended energy range, likely due to hybridization with the rare earth $5d$ orbitals. It should be noted that although the $3d_{3z^2-r^2}$ orbital involvement in the charge order formation is nonzero, its contribution is much less than the hybridized $3d_{x^2-y^2}$ and $2p_{\sigma}$ orbitals, as indicated by the stronger charge order peak in the $\sigma$ polarization channel. These factors mean that minimal theoretical models of charge order in nickelates must explicitly include both Ni and O states alongside strong correlations. Another result of our model is that the doped sites in charge ordered nickelates are much closer to a low-spin $S=0$ state than to a high-spin $S=1$ state, unlike La$_{2-x}$Sr$_x$NiO$_4$, whose high-spin physics drives insulating behavior across the vast majority of its phase diagram \cite{Ulbrich2012neutron}.

Recently, \gls*{RIXS} measurements in infinite-layer nickelate films have discovered and studied charge order at $\mathbf{Q}_{\parallel}=(1/3, 0)$ in undoped and underdoped samples \cite{Rossi2021Broken,Tam2021Charge,Krieger2021Charge}, resembling the charge order in cuprates, but differing from the diagonal charge order in \La438{}. In terms of these differing wavevectors, theoretical model studies in the cuprates have shown that charge order at $(Q,0)$ and $(Q,Q)$ are close in energy, the eventual choice of the charge order wavevector being sensitive to details of the electronic structure and correlations \cite{Melikyan2014Symmetry,Corboz2014Competing}. This idea is supported by the experimental observation that the doping dependent charge order wavevector varies in different cuprate families \cite{Frano2020charge}, similar to what has been seen more recently in the infinite-layer nickelates \cite{Rossi2021Broken}. In view of this, the difference in wavevector probably does not reflect a difference in the mechanisms at play in charge order formation. It should, however, be noted that the parallel charge order seen in infinite-layer materials occurs at a lower hole concentration. 

More information can be obtained by comparing the states involved in charge order formation for different low-valence nickelates \cite{Rossi2021Broken,Tam2021Charge,Krieger2021Charge}. All these recent works support an appreciable role for Ni in charge order formation. However, controversy exists regarding whether the rare-earth-Ni hybridization is crucial for charge order formation \cite{Tam2021Charge}, or whether the charge modulation on rare-earth states only plays a secondary parasitic role \cite{Rossi2021Broken}. Our results support the latter scenario in \La438{}. Regarding the involvement of oxygen states, we provide the first spectroscopic modeling that allows this question to be addressed quantitatively. We deduce a mixed charge-transfer/Mott-Hubbard picture for the charge order and 70\%/30\% split of Ni vs.\ O contributions to the charge modulation. This contradicts some of the previous suggestions for infinite-layer nickelates, which propose a negligible role for oxygen in charge order formation and that in-plane and out-of-plane Ni states contribute roughly equally \cite{Tam2021Charge}. These differences are puzzling considering that different members of the $R_{n+1}$Ni$_{n}$O$_{2n+2}$ family share similar Ni-O bonding, magnetic exchange \cite{Lin2021strong, Lu2021magnetic}, superconducting transition temperatures \cite{Li2019superconductivity, Osada2020superconducting, Zeng2022Superconductivity, Pan2021super}, and calculated electronic structures \footnote{Low valence nickelate electronic structure are rather similar provided they are compared at similar effective dopings \cite{Karp2020comparative, LaBollita2021electronic}}. Part of the challenge of making this comparison is that \gls*{RIXS} maps of infinite-layer films, as well as their charge order properties, vary substantially between different samples of nominally the same composition \cite{Rossi2021Broken,Tam2021Charge,Krieger2021Charge}. In this regard, our quantitative spectroscopic analysis on single crystals is valuable considering that these samples show more consistent spectral properties than films of infinite layer materials \cite{Rossi2021Broken,Tam2021Charge,Krieger2021Charge}.

\section{Conclusion}
In summary, we have used \gls*{RIXS} measurements at the Ni $L_2$-edge to study the character of the electronic structure and charge order in the low-valence nickelate \La438{}. Our work is unique in providing a realistic quantitative empirical model for charge order and validating it using \Q{}-resolved spectroscopy at the charge order wavevector. Different from cuprates where the spatial charge modulation dominantly resides on ligand orbitals, the charge order in \La438{} is mostly contributed by the Ni sites due to the larger charge transfer energy in low-valence nickelates. In addition to the dominant role of in-plane Ni $3d_{x^2-y^2}$ and O $2p_{\sigma}$ orbitals, the out-of-plane Ni $3d_{3z^2-r^2}$ orbitals also participate in the charge order, this being enabled by their hybridization with the rare-earth $5d$ orbitals. Thus, our results reveal that the overall low-energy physical properties of low-valence nickelates are shaped by Ni $3d_{x^2-y^2}$ and O $2p_{\sigma}$ orbitals, while the detailed electronic structure is fine tuned by Ni $3d_{3z^2-r^2}$ and rare-earth $5d$ orbitals. This reveals that multi-orbital physics is crucial to low-valence nickelates, indicating that several different ground states are close in energy. This observation points to a more complex, and perhaps an even richer, phenomenology than their cuprate cousins, while charge order remains an intrinsic character of these strongly correlated materials.

The RIXS data generated in this study have been deposited in the Zenodo database under accession code [to be assigned].

\section*{Acknowledgments}
Work at Brookhaven and the University of Tennessee (RIXS measurements and the interpretation and model Hamiltonian calculations) was supported by the U.S.\ Department of Energy, Office of Science, Office of Basic Energy Sciences, under Award Number DE-SC0022311. Work at Argonne was supported by the U.S. DOE, Office of Science, Basic Energy Sciences, Materials Science and Engineering Division (nickelate sample synthesis and first principles calculations). Work performed at Harvard University (data interpretation and paper writing) was supported by the US Department of Energy, Division of Materials Science, under Contract No.\ DESC0012704.  This research used resources at the SIX beamline of the National Synchrotron Light Source II, a U.S.\ DOE Office of Science User Facility operated for the DOE Office of Science by Brookhaven National Laboratory under Contract No.~DE-SC0012704.

\appendix

\section{\label{sec:synthesis}Sample synthesis}
Parent Ruddlesden-Popper La$_{4}$Ni$_{3}$O$_{10}$ and Pr$_{4}$Ni$_{3}$O$_{10}$ were prepared using the high-pressure optical floating zone method. Sample reduction was performed by cleaving small crystals from the boules and heating them in a flowing H$_2$/Ar gas mixture as described previously \cite{Zhang2017large}. We adopt the tetragonal notation with space group $I4/mmm$ and lattice constants of $a=b=3.97$~\AA{}, $c=26.092$~\AA{} to describe reciprocal space. Using this notation, the samples had a $c$-axis surface normal. The high quality of these samples is confirmed by prior studies \cite{Lin2021strong, Shen2022Role}. Single crystals of \La438{} are particularly suitable for this study as they exhibit more consistent \gls*{XAS} spectra and charge order properties than thin films of infinite-layer nickelates \cite{Rossi2021Broken,Tam2021Charge,Krieger2021Charge}].

\section{\label{sec:RIXS}RIXS measurements}
High-energy-resolution \gls*{RIXS} measurements were performed at the SIX beamline at the NSLS-II. Although the sample geometry and the energy of the Ni $L_2$-edge resonance limits reciprocal space access, charge order in \La438{} has a $c$-axis correlation length of less than one unit cell, which means that the charge order Bragg peaks are accessible for a wide range of $L$ values \cite{Zhang2016stacked}. We chose to measure at the Ni $L_2$-edge instead of the $L_3$ edge to avoid contamination from the La $M$-edge which is very close to the Ni $L_3$-edge and can strongly distort the resonant process \cite{Schussler-Langeheine2005spectroscopy}. In view of this, we fixed the spectrometer angle at its maximum value of $2\Theta=153^{\circ}$ throughout the measurements of the charge order peak. The samples were aligned with the crystalline [0, 0, $L$] and [$H$, $H$, 0] directions lying in the horizontal scattering plane to access the charge order peak with momentum transfer $\mathbf{Q}_{\mathrm{CO}}=(1/3, 1/3, L)$ where $L\approx 1.75$. In this geometry, the x-ray intensity is dominated by charge, rather than spin, scattering (Supplemental Material Sec.~IV \cite{supp}). Spectra designed to study the charge order resonance in the $\sigma$ polarization channel, such as Fig.~\ref{fig:Emap}(e), were taken with 24~meV energy resolution. For the charge order in the $\pi$ polarization channel, such as Fig.~\ref{fig:Emap}(f), a relaxed energy resolution of 32~meV was used to increase throughput. Whenever the energy was changed, the sample was rotated in order to remain at the same in-plane scattering vector. In order to study the high-energy features, as done in Fig.~\ref{fig:Emap}(a) and \ref{fig:Emap}(b), the energy resolution was further relaxed to 48~meV and the sample and spectrometer were slightly offset from the diffraction condition with a sample angle of 14.3$^\circ$ and a spectrometer angle of $2\Theta=147^{\circ}$ to avoid saturating the detector. Note that the strong elastic intensity overwhelms the low-energy inelastic signals such as that from the magnetic excitations studied previously \cite{Lin2021strong}. Data collected with different energy-resolution configurations were normalized by the $dd$ excitations measured with the same sample geometry.

Upon illumination by very strong elastic scattering from charge order, a weak periodic error was identified in the spectrometer grating which created the weak feature in the energy gain side of Fig.~\ref{fig:Emap}(a). This was confirmed by measuring reference elastic scattering.

\section{\label{sec:ED}Exact diagonalization calculations}
The \gls*{RIXS} spectra and \gls*{REXS} responses presented here were calculated using the Kramers-Heisenberg formula in the dipole approximation through the EDRIXS software \cite{EDRIXS, Wang2019EDRIXS}. The eigenstates for the initial/final and intermediate states are obtained from exact diagonalization of a Ni$_3$O$_{10}$ cluster with four holes and open boundary conditions. To fully take into account the many-body and multi-orbital effects, we explicitly include the Coulomb interactions and nearest-neighbor inter-atomic hoppings in our model, and construct the Hamiltonian in hole language. We use the same parameters as those used in the O $K$-edge calculations which are proved to well describe the \gls*{RIXS} data \cite{Shen2022Role}. By doing so, the charge-transfer energy $\Delta$ is set to 5.6~eV and the on-site Coulomb repulsion to 6.5~eV, locating the material in the mixed charge-transfer/Mott-Hubbard regime of the \gls*{ZSA} scheme. We also include the spin-orbit coupling for the Ni $3d$ electrons, which is very small and is expected to play a minimal role. For simplicity, the scattering angle $2\Theta$ is kept at $150^{\circ}$ and the sample angle is fixed to $\theta=15^{\circ}$.

The total \gls*{RIXS} scattering amplitude is calculated via
\begin{equation}
    \mathcal{F}=\sum_i \mathcal{F}_i e^{i\mathbf{Q}\cdot\mathbf{r}_i}
\end{equation}
where $\mathcal{F}_i$ and $\mathbf{r}_i$ are the scattering amplitude and position of each Ni site, respectively. The charge order peak was then calculated by combining the atomic scattering amplitudes with the phases appropriate for tiling the cluster into the NiO$_2$ plane as shown in Fig.~\ref{fig:schematic}(a).

\bibliography{refs}

\end{document}


\title{Supplemental Material: Electronic character of charge order in square planar low valence nickelates}

\author{Y. Shen}\email[]{yshen@bnl.gov}
\author{J. Sears}
\author{G. Fabbris}
\author{J. Li}
\author{J. Pelliciari}
\author{M. Mitrano}
\author{W. He}
\author{Junjie Zhang}
\author{J. F. Mitchell}
\author{V. Bisogni}
\author{M. R. Norman}
\author{S. Johnston}
\author{M. P. M. Dean}\email[]{mdean@bnl.gov}

\date{\today}

\maketitle

\newcommand\tlc[1]{\texorpdfstring{\lowercase{#1}}{#1}}
\renewcommand{\thetable}{S\arabic{table}}  
\renewcommand{\thefigure}{S\arabic{figure}}

\section{Absence of diagonal charge order in P\tlc{r}$_{4}$N\tlc{i}$_{3}$O$_{8}$}

\begin{figure}
\includegraphics{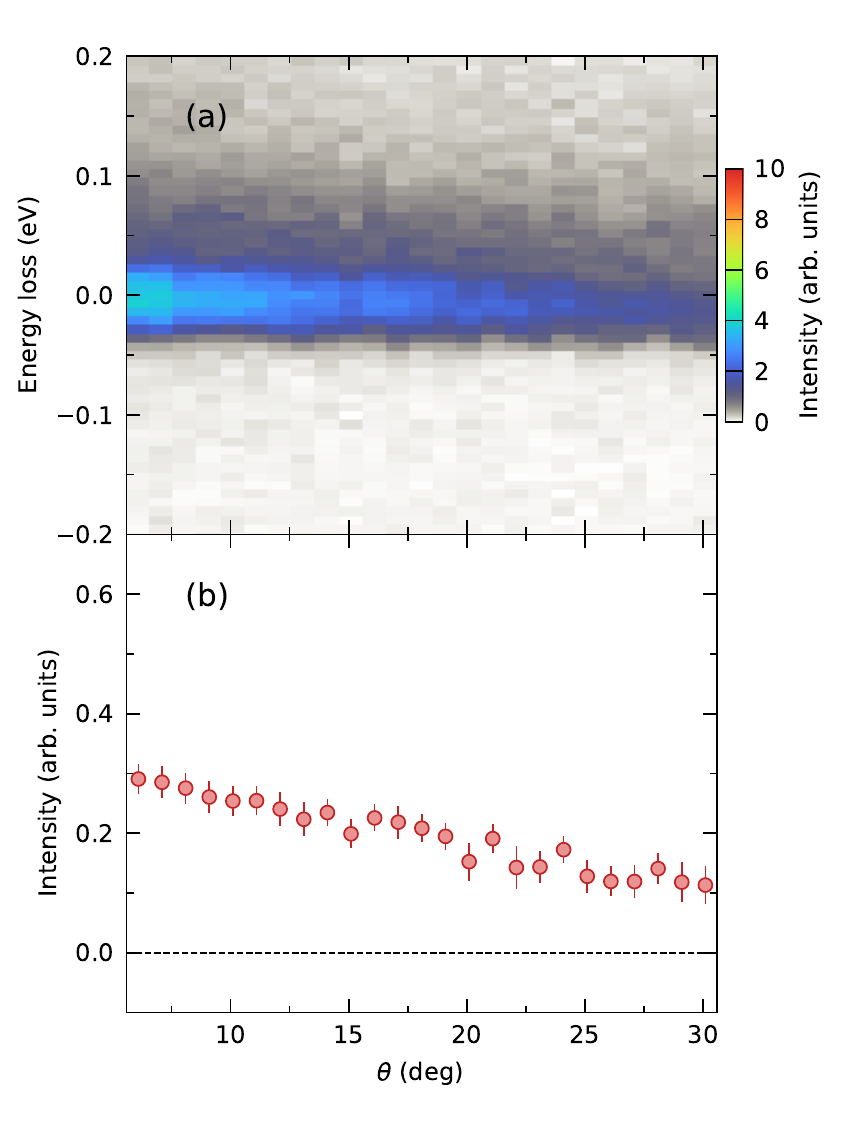}
\caption{Absence of charge order in \Pr438{} at 40~K. (a) \gls*{RIXS} intensity map around $\mathbf{Q}_{\parallel}=(1/3, 1/3)$ in the quasi-elastic regime at the Ni $L_3$-edge. The experimental configuration is the same as that for \La438{}. (b) Quasi-elastic amplitudes extracted from (a).}
\label{fig:PrTdep}
\end{figure}

To confirm the absence of diagonal charge order in metallic \Pr438{} \cite{Zhang2017large}, we performed \gls*{RIXS} measurements near $\mathbf{Q}_{\parallel}=(1/3, 1/3)$ in \Pr438{}. Figure~\ref{fig:PrTdep} shows the \gls*{RIXS} spectra in the quasi-elastic regime with $\sigma$ polarized incident photons. No superlattice peaks are found but only background evolving smoothly with the in-plane sample angle $\theta$, which is primarily caused by the self-absorption effect. Note that despite the absence of long-range or short-range stripe order indicated here, stripe related spin fluctuations are distinguished in the inelastic regime \cite{Lin2021strong}.

\clearpage

\section{RIXS process for different excitations}

\begin{figure*}[!h]
\includegraphics{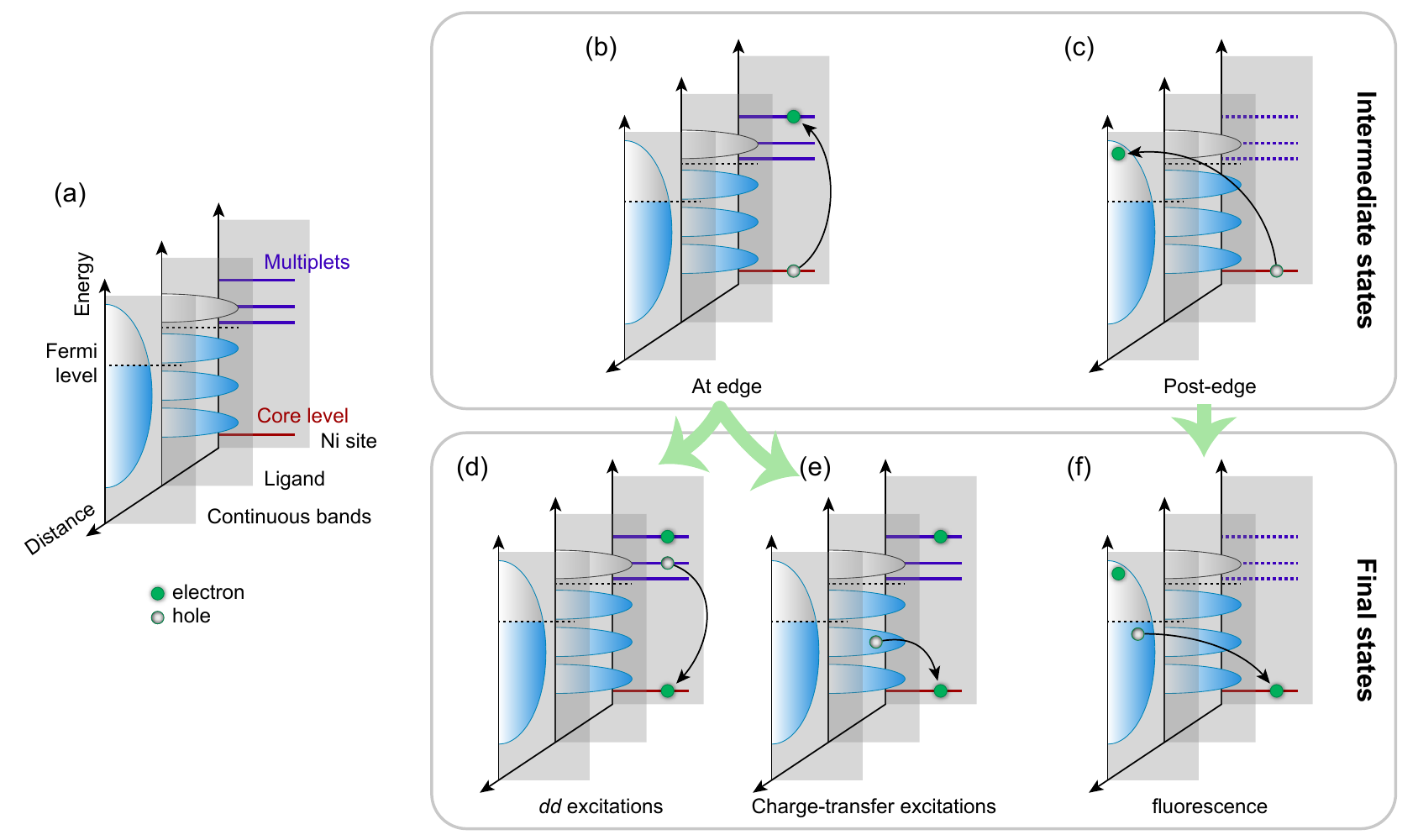}
\caption{RIXS process for different excitations. (a) Legend for each symbol. (b,~c) Photon absorption process and corresponding intermediate states. (d)--(f) Photon emission process and corresponding final states. For the fluorescence excitation scenario, the multiplets are not well defined so they are replaced by dashed lines.}
\label{fig:process}
\end{figure*}

Here we discuss the \gls*{RIXS} process for different excitations. Due to the presence of the strong core-hole potential in the \gls*{RIXS} intermediate states, the electron that is excited from the core level is constrained to a few unit cells near the Ni site where the x-ray absorption takes place. This effect competes with the kinetic energy of the electron and leads to intertwined excitations in the \gls*{RIXS} spectra. In a simplified picture, the orbital states during the \gls*{RIXS} process can be divided into three categories based on how they are affected by the core-hole potential [see Fig.~\ref{fig:process}(a)]. The first one involves the Ni $3d$ orbitals that are strongly localized at the core-hole site. The second one involves the ligand orbitals that surround the Ni site and strongly hybridize with the Ni $3d$ orbitals. They are largely localized but could show a finite bandwidth. The third one involves continuous electronic bands that are mostly unperturbed by the core-hole potential and behave itinerantly with an appreciable bandwidth. The localized Ni $3d$ orbitals can hybridize with the continuous bands in an orbital dependent fashion. At the Ni $L$-edge, the core electron is predominantly excited to the unoccupied localized Ni $3d$ orbitals [see Fig.~\ref{fig:process}(b)]. During the photon emission process, either an electron from another $3d$ orbital deexcites to fill the core hole, leading to $dd$ multiplet excitations [see Fig.~\ref{fig:process}(d)], or an electron from the ligand orbitals hops to the Ni site, resulting in charge-transfer excitations [Fig.~\ref{fig:process}(e)]. Since the ligand orbitals normally lie at a lower energy, the charge-transfer excitations usually occur with a larger energy loss than the $dd$ excitations and are much weaker at the Ni $L$-edge as they are made possible through hybridization. In the post-edge regime, the core electron is excited to the unoccupied states in the continuous bands through hybridization in the intermediate state [see Fig.~\ref{fig:process}(c)], and during the photon emission an electron below the Fermi level deexcites to fill the core hole, leading to the fluorescence [see Fig.~\ref{fig:process}(f)]. As the deexciting process is dominated by electrons near the Fermi energy, fluorescence tends to present a constant emission photon energy. Note that at the Ni $L$-edge \gls*{RIXS} process, contributions from rare earth and oxygen states are seen via their hybridization with atomic Ni $3d$ orbitals. In real materials, there are no clear boundaries between the localized orbitals and continuous bands. Thus, different excitations are also intertwined but the weights are quite different, which helps us distinguish them in the \gls*{RIXS} spectra.

\section{Polarization dependence of fluorescence}

\begin{figure*}
\includegraphics{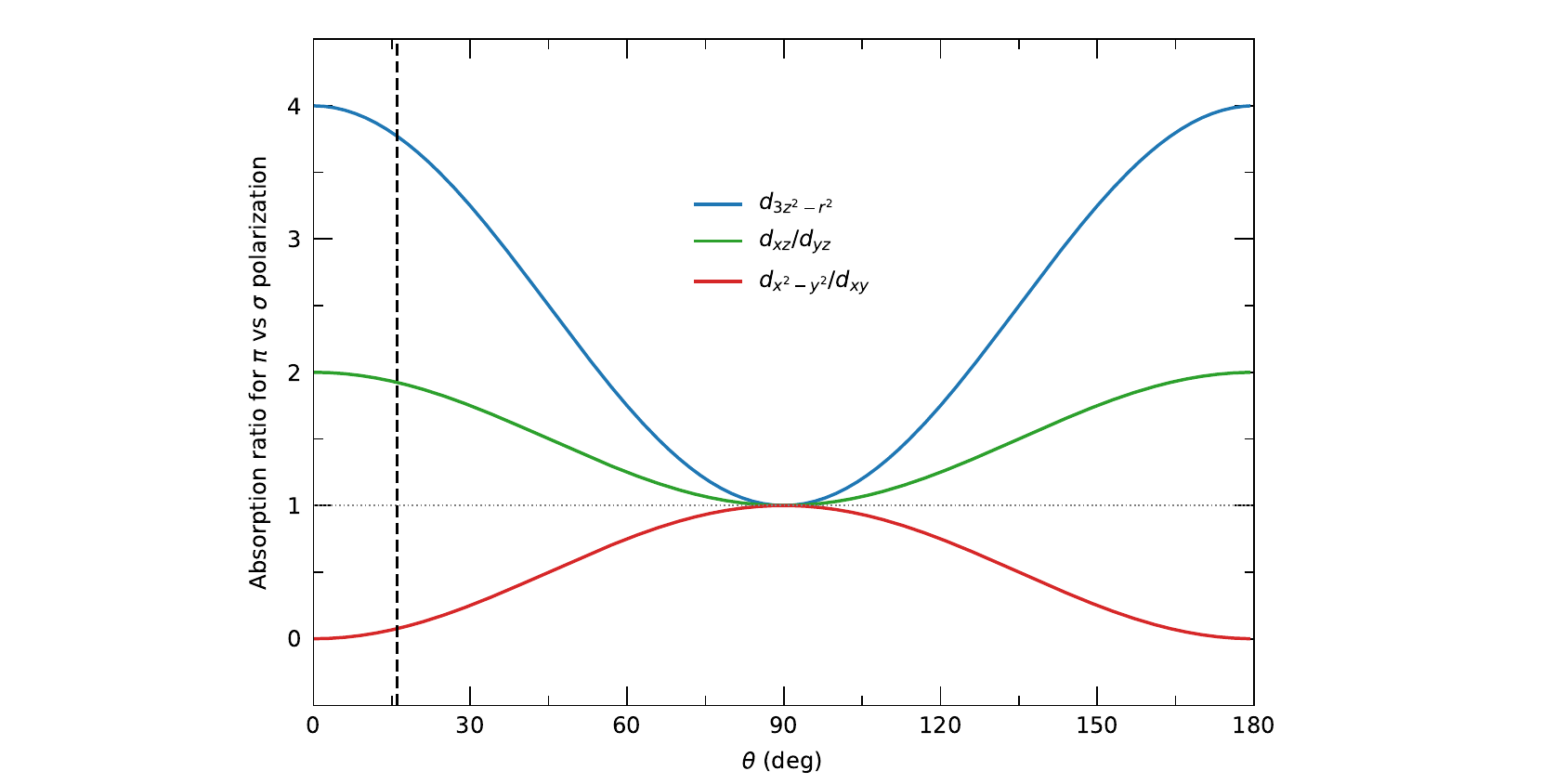}
\caption{Intensity ratio for dipole absorption between $\pi$ and $\sigma$ polarization channels as a function of sample angle $\theta$. The vertical dashed line indicates the experimental configuration we used for charge order measurements while the horizontal dotted line denotes a unit ratio.}
\label{fig:pol_calc}
\end{figure*}

The polarization dependence of $dd$ excitations in cuprates and nickelates has been widely discussed \cite{Rossi2020orbital,Moretti2011energy}. Here, we focus on fluorescence to show how the orbital information can be extracted by comparing the \gls*{RIXS} intensity in different polarization channels. Since we are measuring at the Ni $L$ edge, the \gls*{RIXS} signal can only arise from either Ni orbitals or Ni states hybridized with other orbitals. For the fluorescence features, the photon emission process is quite similar, with electrons from the crystalline environment surrounding the Ni site deexciting to fill the core hole. Hence, the main intensity difference between these two polarization channels comes from the photon absorption process, the cross-section of which can be simulated in the dipole approximation. Fig.~\ref{fig:pol_calc} presents the x-ray dipole absorption intensity ratio between $\pi$ and $\sigma$ polarization channels as a function of sample angle $\theta$. For the experimental configuration we used (vertical dashed line), the biggest contribution to the $\pi$ over $\sigma$ polarization intensity ratio is the Ni $3d_{3z^2-r^2}$ orbitals while $3d_{x^2-y^2}$ and $3d_{xy}$ contribute equally to $\sigma$ over $\pi$ polarization intensity ratio. Since the $3d_{x^2-y^2}$ orbitals dominate near the Fermi level and are expected to show stronger hybridization with oxygen orbitals \cite{Haverkort2012multiplet}, $3d_{xy}$ orbitals are expected to play a less important role. Moreover, the $t_{2g}$ states do not make any significant contribution to the unoccupied states. Thus, we focus on Ni $e_g$ orbitals during the discussion in the main text, which are the subject of most of the debates over the appropriate theoretical models.

\section{Minimal contribution of spin order to the REXS signal}

In \La438{}, spin order takes place concomitantly with the charge order and shares the same $\mathbf{Q}_{\parallel}$. Hence we need to invoke cross-section considerations to separate the possible contribution of charge and spin order \cite{Haverkort2010theory}.

\begin{figure}
\includegraphics{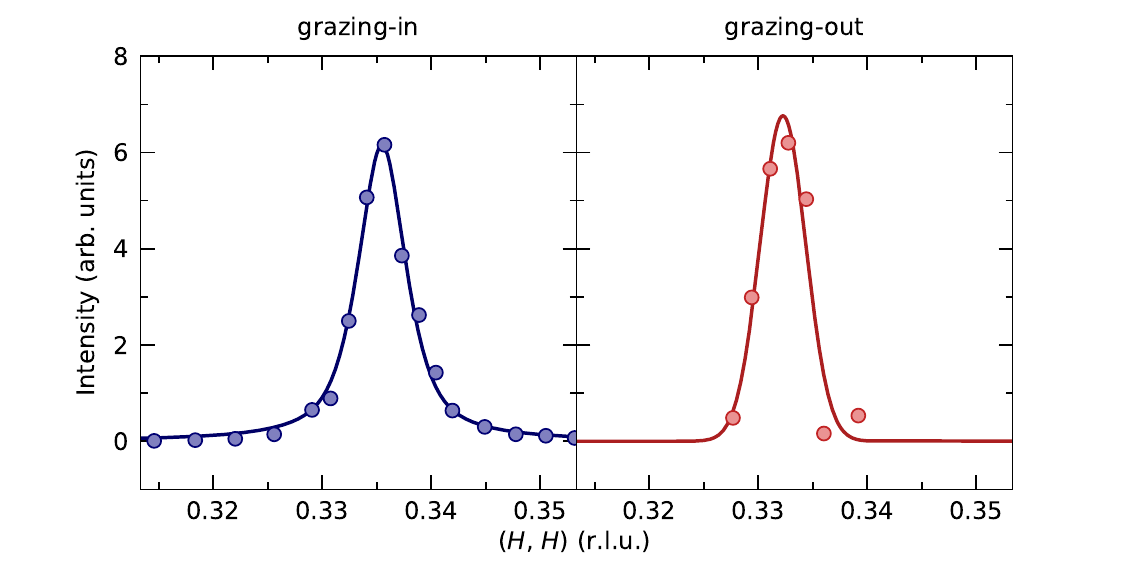}
\caption{Comparison of the superlattice peak intensity with grazing-in and grazing-out conditions. The scattering angle $2\Theta$ was fixed to 153$^{\circ}$ and the data were collected in $\sigma$ polarization channel at 40~K. The solid lines are guides to the eye. Both peaks are found to have essentially the same intensity, which confirms that the peak arises from charge, rather than spin, order.}
\label{fig:gout}
\end{figure}

With $\pi$ incident x-ray polarization, charge order contributes to the measured signal in the $\pi$-$\pi'$ scattering channel while the spin order is responsible for the $\pi$-$\sigma'$ channel. The \gls*{REXS} intensity ratio between these channels can be estimated by $(\bm{k}_i\cdot \bm{k}_f)^2/(\bm{\epsilon}_i\times\bm{\epsilon}'_f\cdot \bm{M})^2=\cos^2{2\Theta}/\sin^2{\theta}\approx 11.8$, where $\bm{k}_i$ ($\bm{k}_f$) is the initial (final) x-ray wavevector, $\bm{\epsilon}_i$ ($\bm{\epsilon}'_f$) is the initial (final) x-ray polarization, and $\bm{M}$ is the spin direction, which is parallel to the $c$-axis in this case. Based on this formula we can see that the \gls*{REXS} signal with $\pi$ incident polarization is dominantly of charge order origin.

Regarding the spin order contribution with $\sigma$ incident x-ray polarization, we can compare the peak intensity with grazing-in and grazing-out conditions. Since the charge order composes the $\sigma$-$\sigma'$ channel, its intensity is expected to be the same in these two geometries. For spin order signal that is only observable in the $\sigma$-$\pi'$ channel, the intensity ratio between grazing-in and grazing-out conditions is $(\bm{k}_{f,~\mathrm{grazing-in}}\cdot \bm{M})^2/(\bm{k}_{f,~\mathrm{grazing-out}}\cdot \bm{M})^2\approx 5.6$, indicating that the spin order signal should be strongly suppressed with grazing-out condition. Figure~\ref{fig:gout} shows the \Q{} dependence of the superlattice peak with both conditions, which are comparable with each other, proving that the superlattice peak observed with $\sigma$ incident x-ray polarization is also dominantly of charge order origin.

\section{Charge order in ED calculations.}

\begin{table*}
\caption{Full list of parameters used for the \gls*{ED} calculations. The on-site orbital energies, hopping integrals and Coulomb interactions are kept the same as those used in the O $K$-edge calculations \cite{Shen2022Role}, and $V_{pd\pi}=-V_{pd\sigma}/2$, $V_{pp\pi}=-V_{pp\sigma}/4$. The potential difference, \DEd{}, only applies to the Ni $3d$ orbitals. Note that the crystal field splitting that is instead used in a Ni atomic model is a combination of point charge potential and orbital hybridization, which can be estimated through ligand field theory \cite{Haverkort2012multiplet}. The resulting effective crystal field splitting gives $10D_q=0.971$, $\Delta e_g=1.041$, $\Delta t_{2g}=0.342$~eV, which are of a similar energy scale as the $dd$ excitations observed in the \gls*{RIXS} measurements. $\zeta_{\mathrm{i}}$ and $\zeta_{\mathrm{n}}$ are spin-orbit coupling parameters of the Ni $3d$ electrons for the initial and intermediate states, respectively, and $\zeta_{\mathrm{c}}$ is the spin-orbit coupling strength for the Ni $2p$ core electrons. The core-hole lifetime is set to be 0.6~eV. All parameters are in units of eV.}
\begin{ruledtabular}
\begin{tabular}{ccccccccc}
\multicolumn{6}{c}{On-site orbital energies} & & \multicolumn{2}{c}{Hopping integrals} \\
$\epsilon_{d_{x^2-y^2}}$ & $\epsilon_{d_{3z^2-r^2}}$ & $\epsilon_{d_{xy}}$ & $\epsilon_{d_{xz/yz}}$ & $\epsilon_{p_{\sigma}}$ & $\epsilon_{p_{\pi}/p_z}$ & & $V_{pd\sigma}$ & $V_{pp\sigma}$ \\
0 & 0.2 & 0.1 & 0.3 & 5.6 & 6.1 & & 1.57 & 0.6 \\
\hline
\multicolumn{3}{c}{Spin-orbit coupling} & & \multicolumn{5}{c}{On-site Coulomb interactions} \\
$\zeta_{\mathrm{i}}$ & $\zeta_{\mathrm{n}}$ & $\zeta_{\mathrm{c}}$ & & $F^0_{dd}$ & $F^2_{dd}$ & $F^4_{dd}$ & $F^0_{pp}$ & $F^2_{pp}$ \\
0.083 & 0.102 & 11.507 & & 5.58 & 6.89 & 4.31 & 3.3 & 5 \\
\hline
\multicolumn{4}{c}{Inter-site Coulomb interactions} & & \multicolumn{4}{c}{Core-hole potential}\\
$F^0_{dp}$ & $F^2_{dp}$ & $G^1_{dp}$ & $G^3_{dp}$ & & $F^0_{dp}$ & $F^2_{dp}$ & $G^1_{dp}$ & $G^3_{dp}$\\ 
1 & 0 & 0 & 0 & & 7.869 & 5.405 & 4.051 & 2.304 \\
\end{tabular}
\end{ruledtabular}
\label{table:allparams}
\end{table*}

We use cluster \gls*{ED} to study the charge order in the low-valence nickelate \La438{}. The full list of the parameters used is presented in Table~\ref{table:allparams}. The validity of our cluster model and parameters has been verified by calculating the \gls*{RIXS} energy maps and confirming that they capture the main features of the measurements as shown in the main text. In the calculations, we include all the Ni $3d$ and O $2p$ orbitals, which leads to a large Hilbert space and correspondingly only a limited number of states can be solved for. Fortunately, the accessible energy range covers the $dd$ excitations so that we can make a direct comparison with the experimental data. The calculated results are broadened using a Gaussian profile with a full width at half maximum of 0.3~eV and are shown in Fig.~3 of the main text.

\begin{figure*}
\includegraphics{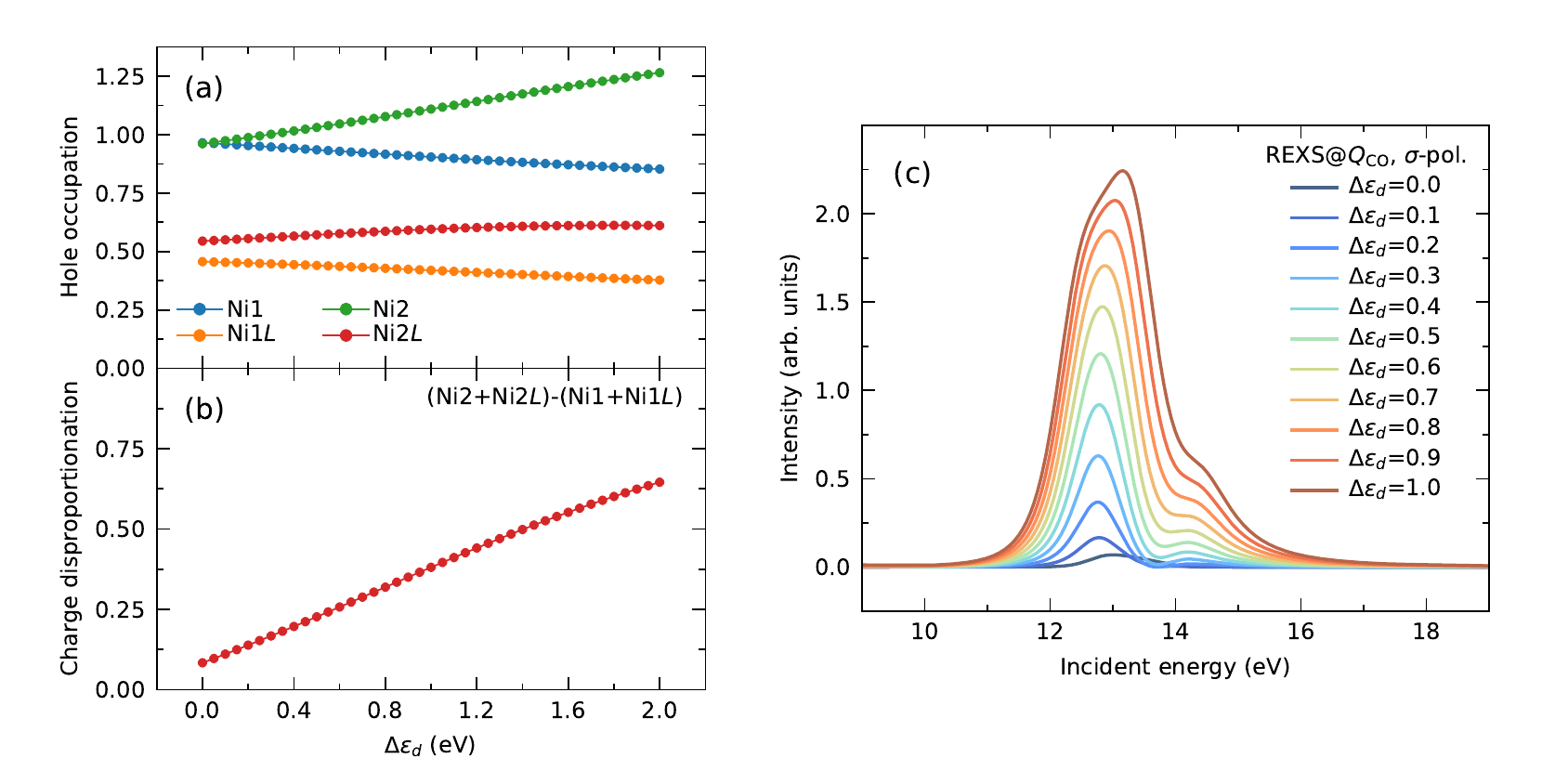}
\caption{The emergence of charge order by introducing the potential difference term \DEd{}. (a) Hole occupations of different sites as a function of \DEd{}. Ni1$L$ stands for the ligand orbitals for Ni1 (the surrounding four oxygens). Correspondingly, one oxygen is shared by Ni1$L$ and Ni2$L$. (b) Charge disproportionation defined as the hole occupation difference of Ni1+Ni1$L$ and Ni2+Ni2$L$. (c) Calculated \gls*{REXS} signals at \QCO{} with different \DEd{}. All the calculations are performed with $\Delta=5.6$~eV.}
\label{fig:DEd_dep}
\end{figure*}

To fully explore the charge order character in the \gls*{ED} calculations, we need to cover a large incident energy range but only the ground state is needed to calculate the \gls*{REXS} signals. Thus, we only include the Ni $3d_{x^2-y^2}$ and O $2p_{\sigma}$ orbitals during the calculations of the charge order, which dominate the ground state, so that a tractable basis size is realized. To trigger charge order in the Ni$_3$O$_{10}$ cluster, we introduce a potential difference \DEd{} as described in the main text. In a microscopic model like we use here, the onsite energy shift and charge occupation are intrinsically coupled, which is different from a phenomenological model where these two factors can be tuned independently \cite{Achkar2012Distinct,Achkar2013resonant}. As shown in Fig.~\ref{fig:DEd_dep}, when \DEd{} is zero, the hole occupations on different Ni sites are almost the same while the hole occupations of ligand orbitals are slightly imbalanced since Ni2 shares oxygens with both Ni1 and Ni3, leading to a small charge disproportionation. With increasing \DEd{}, the charge imbalances on both the Ni and ligand orbitals are enhanced with the former much more prominent, indicating that most of the spatial charge modulation resides on the Ni sites, leading to a Ni site-centered charge order. Correspondingly, a charge-order peak emerge in the \gls*{REXS} calculations, the intensity of which increases with increasing charge disproportionation while the lineshape only evolves by a little.

\begin{figure*}
\includegraphics{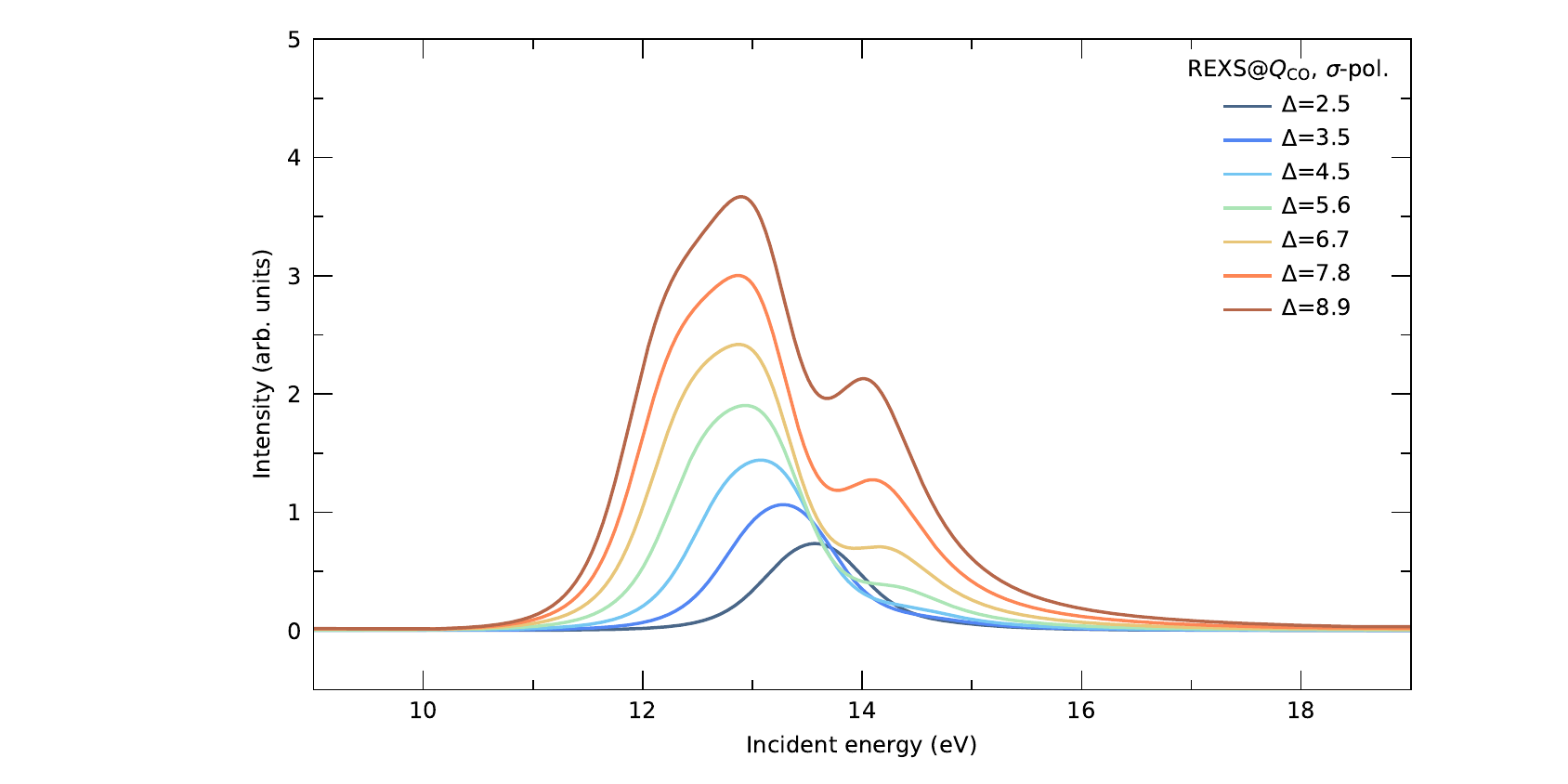}
\caption{Calculated \gls*{REXS} signals at the charge order wavevector \QCO{} with different charge-transfer energy $\Delta$. All the calculations are performed with $\Delta\epsilon_d=0.8$~eV and $U=6.5$~eV.}
\label{fig:delta_dep}
\end{figure*}

After testing the effect of \DEd{}, here we compare results with different charge-transfer energy $\Delta$ in addition to the calculated results presented in the main text. As shown in Fig.~\ref{fig:delta_dep}, in the charge-transfer regime ($\Delta\ll U_{dd}$), a sharp resonant peak is obtained, resembling the experimental observations in cuprates. With increasing $\Delta$, the \gls*{REXS} lineshape evolves correspondingly. In the Mott-Hubbard limit ($\Delta\gg U_{dd}$), the charge order peak becomes broader and shows multiple peak features. Compared with the data presented in the main text, we conclude that a charge-transfer energy with an intermediate strength ($\Delta\approx U_{dd}$) matches the experimental results the best.

\bibliography{refs}